# Efficient, inverse large-scale optimization of diffractive lenses

MARCO GERHARDT,[1,2] SUNGKUN HONG,[1,2] AND MOOSUNG LEE[1,2,*]

[1]Institute for Functional Matter and Quantum Technologies, University of Stuttgart, 70569 Stuttgart, Germany

[2]Center for Integrated Quantum Science and Technology, University of Stuttgart, 70569 Stuttgart, Germany

*moosung.lee@fmq.uni-stuttgart.de

**Abstract:** Scalable photonic optimization holds the promise of significantly enhancing the performance of diffractive lenses across a wide range of photonic applications. However, the high computational cost of conventional full three-dimensional electromagnetic solvers has thus far been a major obstacle to large-scale-domain optimization. Here, we address this limitation by integrating the convergent Born series with the adjoint-field optimization framework, enabling inverse design with its domain size up to a 110 × 110 × 46 μm$^3$ volume—corresponding to 0.1 gigavoxels—using a single, cost-effective graphics card. The optimized lens achieves a 9% improvement in axial resolution and a 20% increase in focusing efficiency compared to a standard Fresnel lens of identical diameter and numerical aperture. These gains point to application opportunities for optimizing high-performance microscopy, photolithography, and optical trapping systems using modest computational resources.

## 1. Introduction

Modern nanolithography techniques have opened new frontiers in the design of high-performance nanophotonic elements. Among these are lenses that support both high numerical aperture (NA) and focusing efficiency ($\eta_f$), enabling compact, high-resolution optical systems for a wide range of applications, including microscopy [1], holography [2], and laser writing [3].

High-NA, high-$\eta_f$ lenses are particularly critical for optical trapping, in which tightly focused light is used to manipulate microscopic objects ranging from biomolecules and living cells [4] to ultracold atoms [5]. Notably, recent demonstrations of quantum-limited motional control of optically levitated macroscopic objects in vacuum at room temperature [6,7] highlight the growing importance of optimized lens systems in quantum levitated optomechanics [8]. In response, prior studies have explored replacing conventional bulky objective lenses with compact, customized alternatives such as high-NA metalenses [9] and laser-printed Fresnel lenses integrated on fiber tips [10,11].

Recent advances in large-area photolithography now allow the fabrication of high-NA flat optics beyond a few tens of micrometers—for example, 100-μm-diameter metalenses operating at visible wavelengths [12]. These devices provide wide field of view and high resolution, yet conventional metalenses [12,13] and Fresnel lenses [14] still suffer from significant efficiency losses caused by near-field coupling in their sub-wavelength features, limiting $\eta_f$ well below the theoretical limit. Addressing these losses requires optimizing lens structures at fine resolution over macroscopic scales, highlighting the need for a scalable optimization framework.

Various inverse design methods have been proposed, including genetic algorithms, deep neural networks, and objective-first methods [15,16]. However, all face severe computational bottlenecks associated with full-wave electromagnetic simulations. For example, the widely used finite-difference time-domain (FDTD) method scales poorly with voxel count for a standard full-3D simulation, making large-scale optimization for designs exceeding 100 μm laterally computationally costly on a standard computing environment and often requiring expensive resources such as high-end workstations [13] or supercomputers [17].

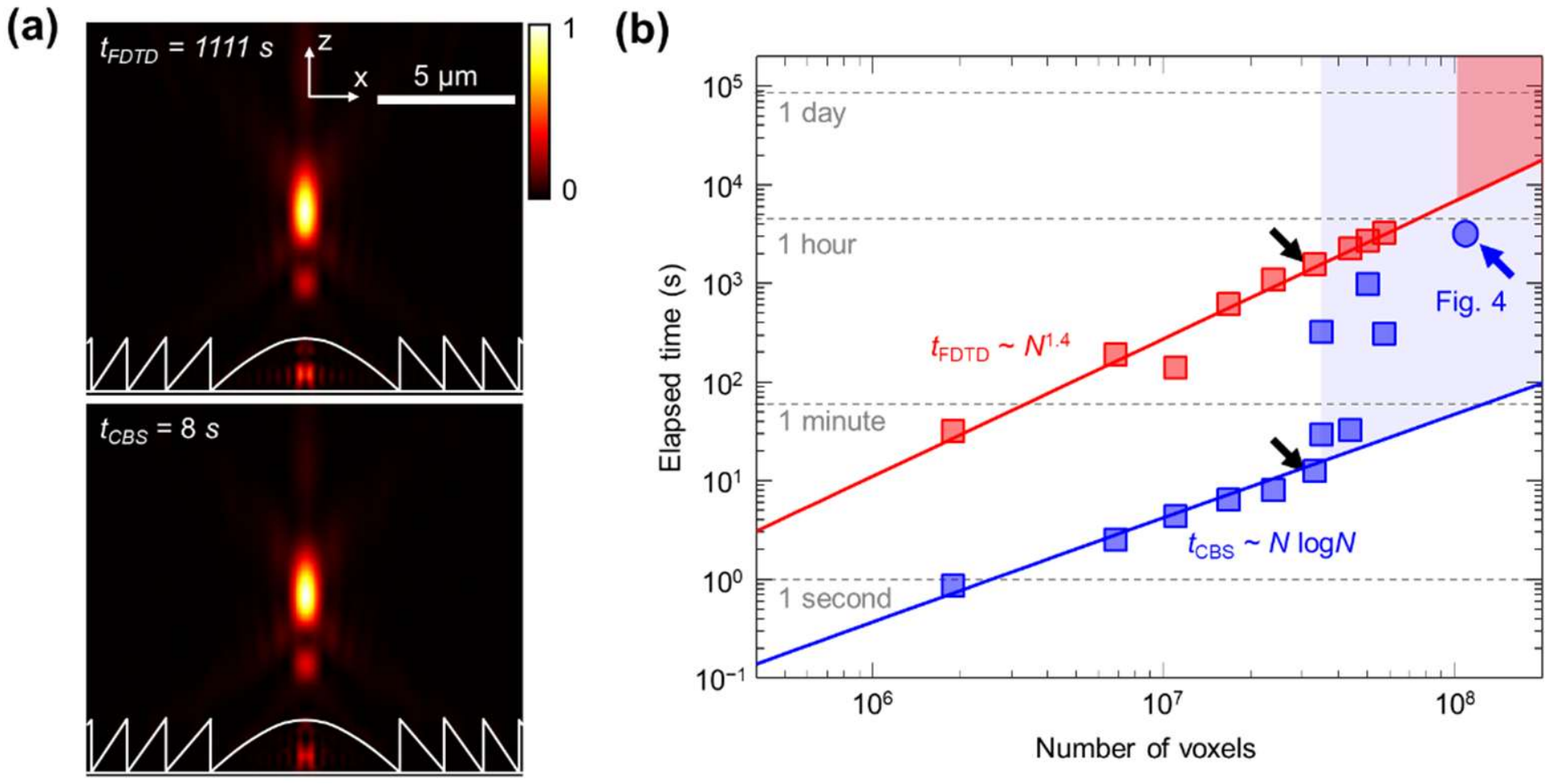

Fig. 1. Comparison of 3D electromagnetic simulation performance. (a) Cross-sectional views (XZ-plane) of the total electric field intensity resulting from diffraction by a Fresnel lens with NA = 0.84 and diameter $D$ = 16 µm. Simulations are performed over a 275 × 275 × 315 grid with spatial resolution λ/16, using the finite-difference time-domain (FDTD, top) and convergent Born series (CBS, bottom) methods. (b) Elapsed computation time for FDTD (red squares) and CBS (blue squares) as functions of the total number of voxels ($N$). Simulations are run under identical conditions with varying voxel sizes. Square points are obtained by varying voxel pitches ranging from λ/6 ($N$ ~ 2 megavoxels) to λ/22 ($N$ ~ 57.2 megavoxels). Power-law fits show scaling of $t_{FDTD} \sim N^{1.4}$ and $t_{CBS} \sim N\log N$, respectively. Black arrows indicate the data points used in (a). The blue shaded area denotes a performance cliff of our method ($N$ > 33 million voxels), where the elapsed computation time diverges from the $t_{CBS} \sim N\log N$ fitting due to the GPU overhead issue. The red shaded area denotes a size limit of FDTD due to the 16 GB memory limit. The blue circle denotes the voxel number of the data used for optimization in Fig. 4.

Here, we address this large-scale lens design problem by integrating the convergent Born series (CBS) with an adjoint-field gradient optimization method. This approach accelerates 3D simulation and inverse design by two orders of magnitude compared to conventional full-3D FDTD solvers until reaching a hardware bottleneck. Using this method, we optimize a large-scale diffractive flat lens in a simulation domain up to a 110 × 110 × 46 µm$^3$ volume (621 × 621 × 261 voxels, 177 nm pitch), totaling 0.1 gigavoxels using a commercial, affordable graphics card. The resulting design surpasses a conventional Fresnel lens in both axial resolution and focusing intensity, demonstrating the feasibility of high-performance, large-scale photonic design on consumer-grade hardware.

## 2. Principle

We begin by evaluating the computational performance of the CBS method, a recently developed algorithm for efficient monochromatic linear electromagnetic simulation [18,19], also used in inverse problems such as tomographic reconstruction [20,21] and EUV mask design [22]. The CBS is a technique that rigorously solves the full-3D inhomogeneous electromagnetic wave equation by converting a Born series to a convergent form [18,19]. A detailed mathematical representation is described in Supplement 1.

To compare the computational performance of the CBS with a conventional full-3D FDTD solver, we simulate light focused by a Fresnel lens with diameter $D$ = 16 µm and NA = 0.84 at wavelength λ = 1064 nm [Fig. 1]. To reduce simulation errors from staircase artefacts at the discrete lens boundaries, subpixel averaging is applied [19]. All tests are run on a consumer-grade desktop PC equipped with an Intel i7-13700K CPU running an open-source Python FDTD package [23], and an Nvidia GeForce RTX 4070 Ti GPU executing a MATLAB 2024b

CBS implementation. Both FDTD and CBS use a perfectly matching absorbing layer of thickness [λ, λ, 6λ] in 3D [24].

At fine voxel resolution (λ/16 pitch, corresponding to 23.8 megavoxels), both methods produce virtually identical intensity distributions [Fig. 1(a)]. Under these conditions, the FDTD simulation requires over 1,000 seconds to complete, whereas CBS completes the task in under 10 seconds. To characterize how computational time scales with problem size, we repeat the simulations while varying the voxel pitch, thereby adjusting the total voxel count under identical physical and hardware conditions [Fig. 1(b)]. The performance of both FDTD and CBS is constrained by memory overhead. For FDTD, simulations beyond 0.1 gigavoxel could not be performed due to the 16 GB memory limit. In contrast, our method was able to carry out simulations beyond this size limit with a smaller VRAM capacity of 12 GB, albeit with significant GPU memory overhead. Specifically, power-law fits reveal that CBS maintains a speed advantage of approximately two orders of magnitude over FDTD across problem sizes up to 33 million voxels. However, once the problem size exceeds this threshold, the runtime steeply increases by more than 10-fold due to significant memory overhead.

For typical research workflows, it is therefore crucial to identify the problem sizes at which the GPU performance bottlenecks occur, in order to fully exploit the acceleration capabilities of our method. For instance, upgrading the current GPU with 12 GB memory to a high-end resource (e.g., NVIDIA RTX Pro 6000 Blackwell, 96GB) is expected to enable high-speed computation for approximately 0.26 gigavoxels without bottleneck.

Building on the efficiency of the CBS solver, we implement a gradient-based inverse-design algorithm to optimize the diffractive lens geometry for improved focal intensity [Fig. 2; see Supplement 1 for mathematical formulation]. Our approach is based on the adjoint method with topology optimization, a well-established framework in photonics [16,25]. Specifically, we treat the lens thickness map $h(\mathbf{r}_\perp)$ as the design variables [Fig. 2(a)]. This choice is directly compatible with two-photon polymerization techniques such as IP-Dip lithography [1]. We initialize $h(\mathbf{r}_\perp)$ as a standard Fresnel lens with a maximum thickness of 2 μm and NA = 0.84, closely matching our prior experiments [11].

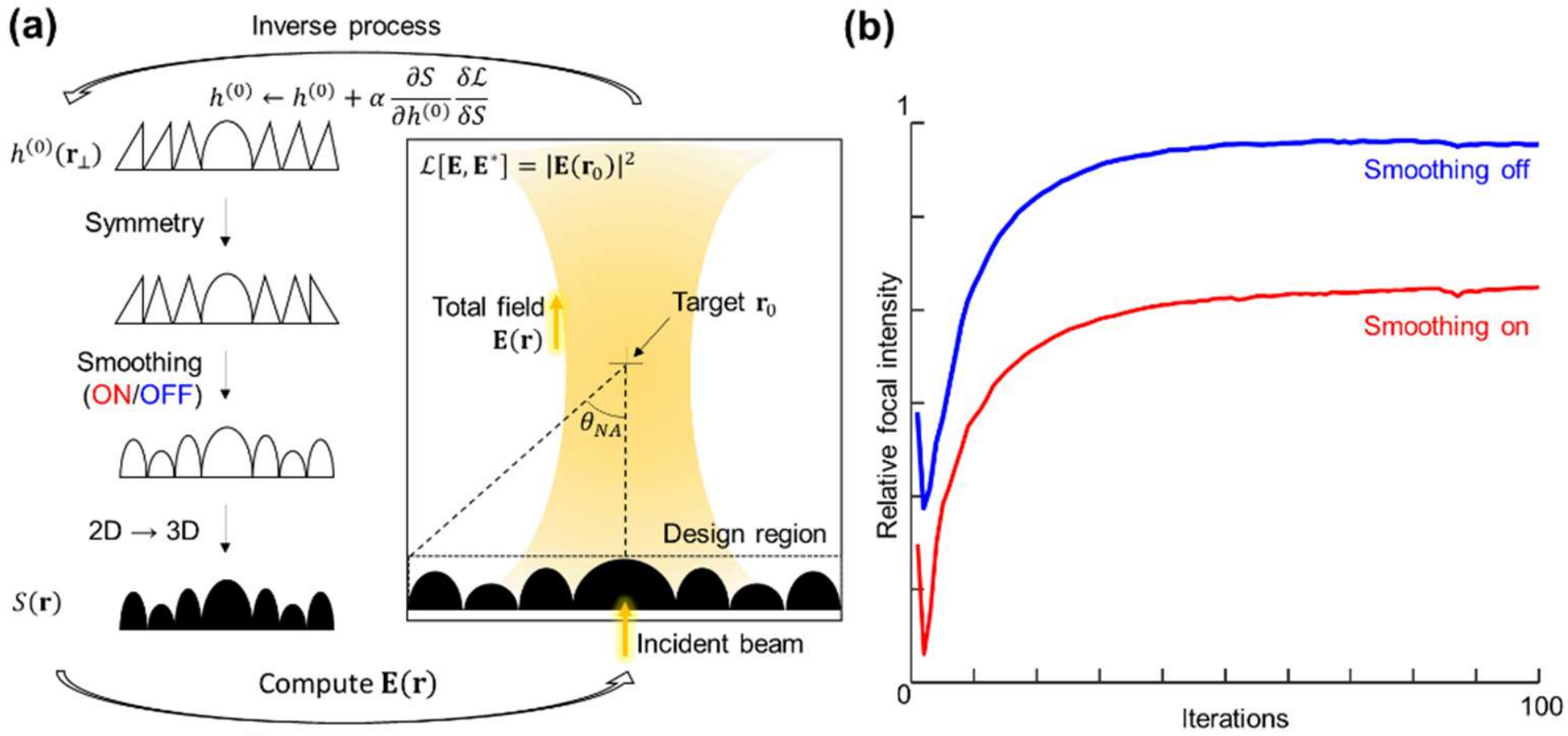

Fig. 2. (a) Schematic illustration of the inverse design process. Left: The lens thickness profile $h(\mathbf{r}_\perp)$ is subject to mirror symmetry and optional Gaussian smoothing, then mapped onto a 3D binary shape function $S(\mathbf{r})$. Right: In the forward step, the structure $S(\mathbf{r})$ is used to compute the total electric field $\mathbf{E}(\mathbf{r})$, from which the figure of merit $\mathcal{L}[\mathbf{E}, \mathbf{E}^*] = |\mathbf{E}(\mathbf{r}_0)|^2$ is evaluated at the target focal point, $\mathbf{r}_0 = (0, 0, f_{eff})$, where $f_{\mathrm{eff}}$ satisfies $\theta_{\mathrm{NA}} = \sin^{-1}(\mathrm{NA}) = \tan^{-1}(D/2f_{\mathrm{eff}})$. The resulting field also provides the gradient used to update the height profile in the inverse design loop. (b) Evolution of relative focal intensity over 100 optimization iterations with (red) and without (blue) Gaussian smoothing.

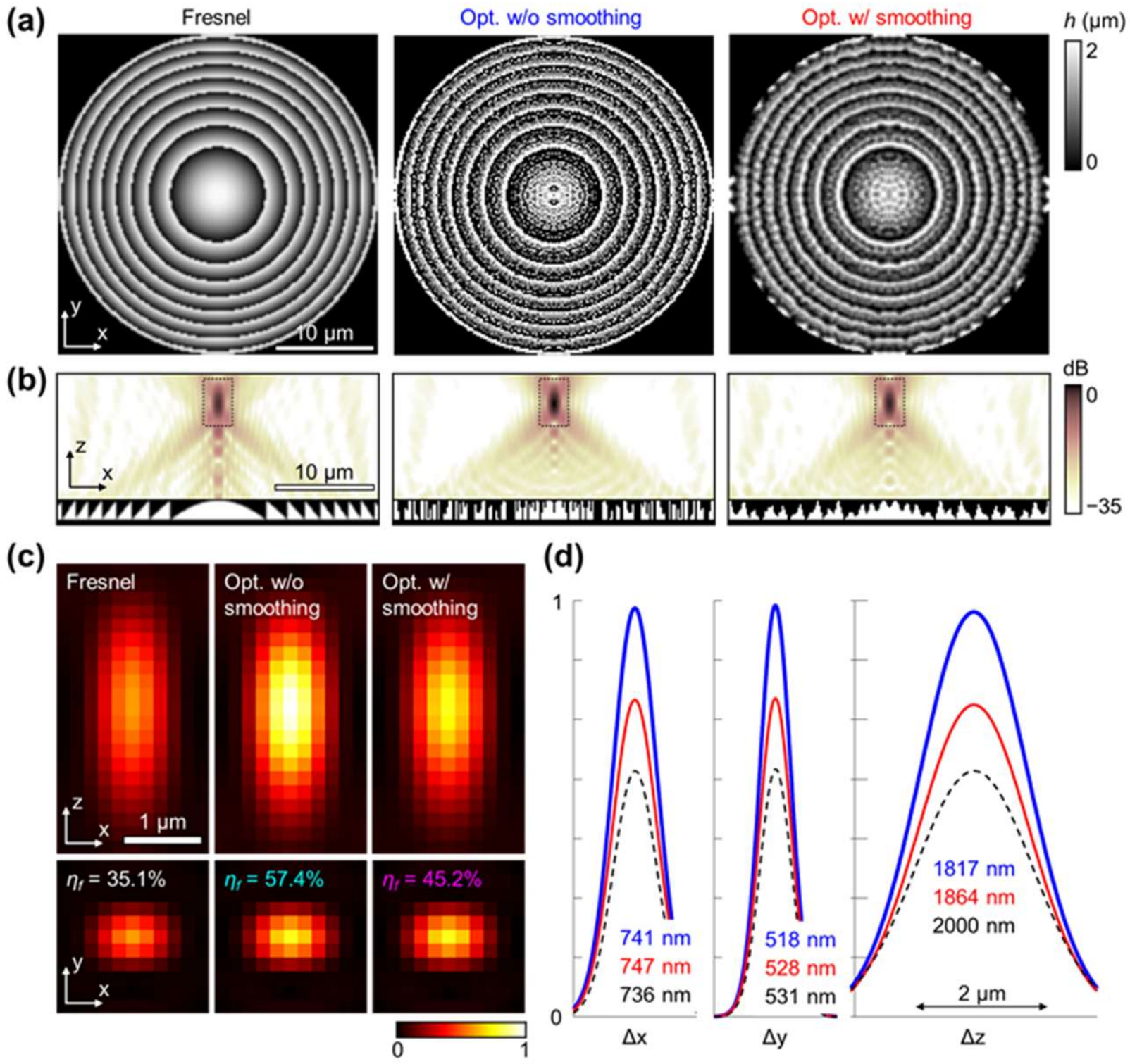

Fig. 3. Results of inverse-designed diffractive lenses with diameter $D$ = 32 μm. (a) Thickness maps of a conventional Fresnel lens (left) and optimized lenses without (middle) and with (right) smoothing constraints (right). (b) Cross-sectional (XZ-plane) views of the total electric field intensity in decibel scale for the Fresnel (left) and optimized lenses without (middle) and with (right) smoothing. (c) Magnified cross-sectional views (XZ and XY planes) of the resulting PSFs, comparing Fresnel and optimized lenses. (d) Relative intensity line plots of the PSFs along the $x$-, $y$-, and $z$-axes. Solid blue and red lines correspond to optimized lenses without and with smoothing constraints, respectively; black dashed lines represent the Fresnel lens. Full widths at half-maximum are labeled.

We incorporate geometric constraints into our topology optimization based on several practical considerations. First, to enforce symmetry of the resulting point spread function (PSF), we impose mirror symmetry on the thickness map, $h(\mathbf{r}_\perp) = h(\pm x, \pm y)$. Second, to account for the minimum feature size achievable with two-photon photolithography [3], we apply Gaussian smoothing with a standard deviation of 355 nm. To assess the effect of this fabrication limit on focusing efficiency, we perform the optimization both with and without the smoothing filter for comparison. The subsequent profile is then mapped onto the 3D shape function $S(\mathbf{r}) = \Theta(z(h(\mathbf{r}_\perp) - z))$, where $\Theta(x)$ is a Heaviside step function. The refractive index of the polymerized lens is set to $n = 1.54$ [26]. Under normal incidence of an $x$-polarized plane wave at $\lambda = 1064$ nm, the CBS computes the total electric field $\mathbf{E}(\mathbf{r})$.

The figure of merit (FoM) is the focal intensity at the target focal position, $\mathbf{r}_0 = (0, 0, f_{eff})$, evaluated as $\mathcal{L}[\mathbf{E}, \mathbf{E}^*] = |\mathbf{E}(\mathbf{r}_0)|^2$. Here, the effective focal length $f_{\text{eff}}$ follows from the relation with the half-angle of the focusing light cone, $\theta_{\text{NA}} = \sin^{-1}(\text{NA}) = \tan^{-1}(\text{D}/2f_{\text{eff}})$; thus, larger diameters yield longer working distances. The 3D gradient $\delta\mathcal{L}/\delta S$ is computed via the CBS, and the derivative $\delta\mathcal{L}/\delta h$ is then obtained through chain rule differentiation, updating the thickness profile to complete one optimization step.

To accelerate convergence, we use the adaptive moment estimation (Adam) optimizer [27], with step size $\alpha = 0.3$, decay rates $\beta_1 = 0.9$, $\beta_2 = 0.999$, and regularization parameter $\varepsilon = 10^{-6}$ [Fig. 2(b)]. Except during the initial step, optimization runs—regardless of smoothing—show steady improvements in focal intensity across 100 iterations. The smoothing constraint reduces the performance gain, consistent with prior results in metalens optimization [13].

## 3. Result

To benchmark the optimization performance, we first evaluate the inverse-designed diffractive lenses with diameter $D$ = 32 μm [Fig. 3]. The simulation domain spans a volume of 4.3-megavoxels (189 × 189 × 121 voxels, 177 nm pitch, 34 × 34 × 21 $\mu m^3$). On a single desktop GPU, the full optimization completes in approximately 370 seconds of wall-clock time, averaging 3.7 seconds per iteration—roughly twice the time required for a single forward calculation.

The resulting lens profiles display intricate sub-wavelength features, particularly in the absence of smoothing constraints [Fig. 3(a)]. In the XZ field profiles, the Fresnel lens shows pronounced diffraction sidelobes arising from its discontinuous structure. In contrast, both optimized designs significantly suppress these sidelobes [Fig. 3(b)]. Consequently, the magnified PSFs reveal a notable increase in focal intensity for the optimized lenses, with the design without smoothing achieving the highest focusing efficiency [Fig. 3(c)].

We quantify the corresponding focusing efficiency $\eta_f$, as the ratio of the total laser power within a circular area of diameter equal to three times the lateral full-width half-maxima at the focal plane to the incident laser power [13]. For a Fresnel lens, the simulation obtains $\eta_f$ = 35.1%. On the other hand, the optimized lens without smoothing yields $\eta_f$ = 57.4%, corresponding to a 1.6-fold improvement and only 9% below the theoretical maximum limit of 63% [13]. When smoothing constraints are applied, the optimized lens yields $\eta_f$ = 45.2%, representing a 1.3-fold enhancement over the Fresnel baseline. Line-plot analysis further confirms a 10% improvement in axial resolution for both optimized designs, regardless of the use of smoothing constraints [Fig. 3(d)].

Having established that smoothing constraints can sufficiently improve $\eta_f$ with manufacturable structures, we perform large-scale optimization with smoothing constraints over a 0.1 gigavoxel simulation volume [Fig. 4]. This voxel count corresponds to 621 × 621 × 261 voxels, with a 177 nm pitch, covering a 110 × 110 × 46 $\mu m^3$ volume [Fig. 4(a)]. The design degrees of freedom, or the number of independent physical design variables, is approximately 300,000. Given this large simulation size, we could not perform FDTD simulation for appropriate simulation due to limited memory capacity and coarse grid resolution (λ/6) [23]. However, our method operates on a single desktop GPU, requiring a total of 7.5 days of wall-clock time, corresponding to 1.8 hours per iteration. The resulting improvement in focal intensity, $\eta_f$ (from 36.5% to 44.2%), and axial resolution (from 2 μm to 1.8 μm) are consistent with those of a smaller diameter shown in Fig. 3 [Fig. 4(b)]. The feature size and lens dimension are fully compatible with state-of-the-art nanofabrication techniques based on two-photon polymerization [1,11].

## 4. Discussion

We introduce a CBS-based adjoint optimization framework as a computationally efficient alternative to conventional FDTD-based approaches. We choose Fresnel lenses as a widely recognized baseline to clearly demonstrate how our method efficiently attains high-performance solutions. The improved $\eta_f$ and spatial resolution demonstrated by the optimized lens imply a corresponding enhancement in light focusing and collection. This is a particularly important factor in recent advances in optical trapping, where enhanced trapping and detection efficiency have facilitated quantum-limited control of optically trapped nanoparticles in a high vacuum [7,8,28,29]. Several prior studies have shown the potential of utilizing metalenses [9] and laser-printed Fresnel lenses [11] in the field of levitated optomechanics. A natural next challenge is to employ inverse design in this context to further optimize both trapping and detection efficiencies.

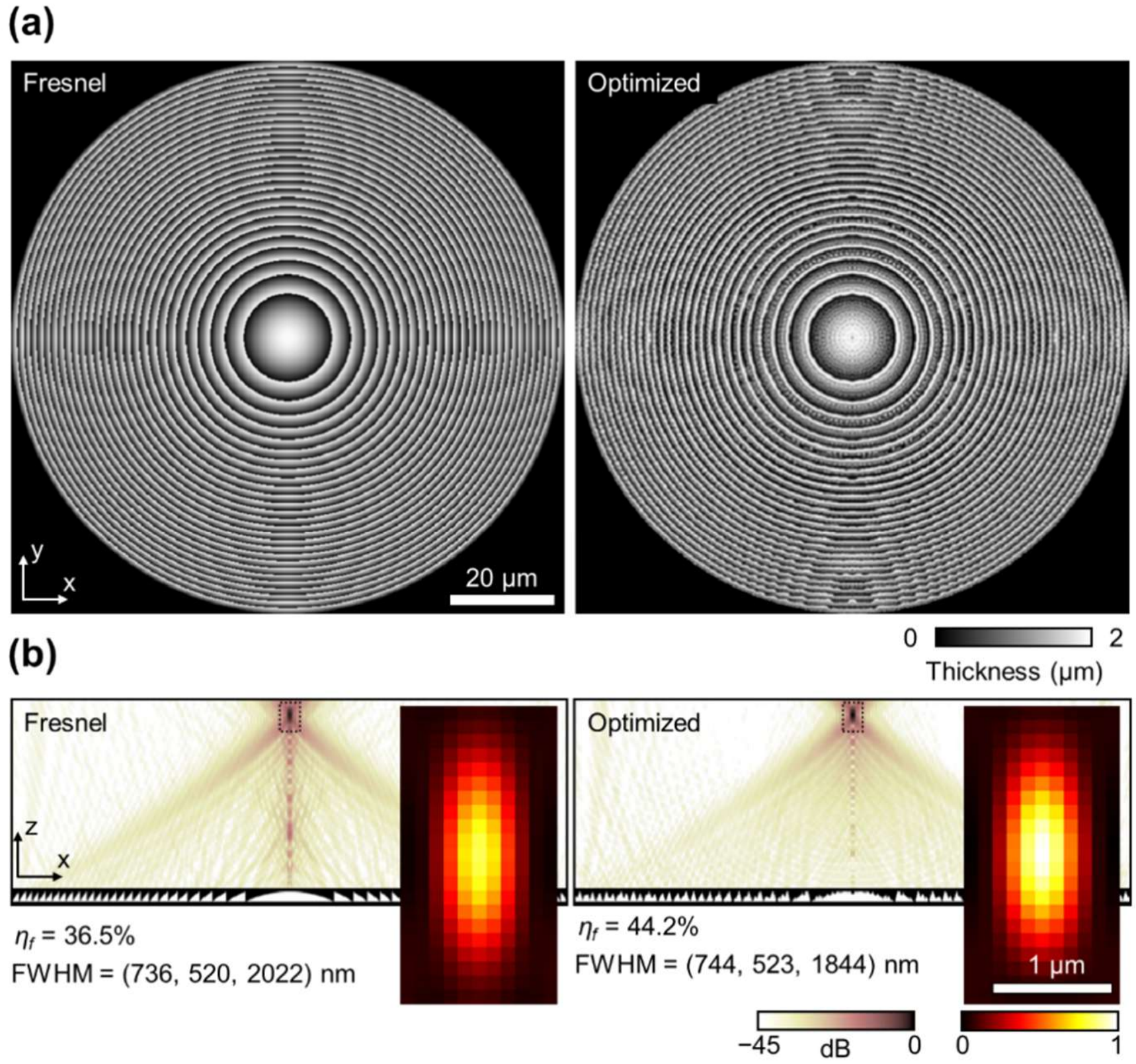

Fig. 4. Large-scale optimization with smoothing constraint. (a) Thickness maps of a conventional Fresnel lens (left) and the optimized lens (right), both with diameter $D$ = 109 μm and NA = 0.84. (b) Cross-sectional views (XZ-plane) of the total electric field intensity in decibel scale for the Fresnel (left) and optimized (right) lenses. Insets: Magnified XZ cross-sectional views of the resulting PSFs.

To predict the expected enhancement in frequency, we utilize our previously developed optical trapping model in the Rayleigh regime, where the particle size is much smaller than the optical wavelength [30] (see Supplement 1 for the detailed derivation). Considering an incident power of $P_0$ = 400 mW, $n$ = 1.45 and $\rho$ = 1850 kg/m$^3$ for a silica nanoparticle, the estimated 3D angular trap frequencies are $[\Omega_x, \Omega_y, \Omega_z] = 2\pi \times [467, 661, 189]$ kHz for the optimized lens and $2\pi \times [413, 562, 153]$ kHz for the Fresnel lens. This corresponds to an enhancement in trap frequencies of 13%, 18%, and 24% along the $x$, $y$, and $z$ axes, respectively.

Notably, despite the optimized structure's lower focusing efficiency ($\eta_f$ = 44.2%) compared to the ideal limit ($\eta_f$ = 63%), its angular trap frequencies are comparable to an ideal Airy disk: $[\Omega_x, \Omega_y, \Omega_z] = 2\pi \times [477, 589, 189]$ kHz. This is attributed to the super-oscillatory lens effect of the optimized lens, which creates a sub-Rayleigh focal spot with 3D FWHMs smaller than those of an ideal Airy disk ([790, 640, 2001] nm at NA = 0.84), albeit at the expense of larger sidelobes [31]. These results highlight a substantial advantage of inverse design in optimizing specific performance metrics in optical trapping applications.

The full capabilities of our method are currently limited to 0.1-gigavoxel problem sizes at the expense of significant GPU memory limit and subsequent overhead issues. For instance, optimizing the 109-μm-diameter lens [Fig. 4] on a single desktop GPU required 1.8 hours per iteration. As shown in Fig. 1(b), the computation time per iteration is more than twice the extrapolated runtime of a forward calculation, as the data size close to the memory limit (12

GB) leads to significant overhead issues. To extend our method's capabilities, an important next challenge is to benchmark the performance against existing state-of-the-art approaches that utilize geometric constraints and dedicated high-performance computing resources. For example, several state-of-the-art studies have successfully scaled inverse design to larger areas by combining FDTD with effective techniques, such as locally periodic approximations [32,33], or imposed symmetry constraints [34]. These techniques are often combined with well-known near-to-far-field transformation techniques to efficiently maximize the design degrees of freedom [35]. A state-of-the-art achieves a gigavoxel-scale optimization through high-end computing resources (e.g., 8 Nvidia A100 GPUs) combined with zoned discrete axisymmetry [36]. We anticipate that these techniques can be complementarily integrated with our method to effectively circumvent the GPU overhead and further enhance scalability.

Our forward solver, the CBS method, is not a simple approximation; it rigorously solves the full 3D inhomogeneous electromagnetic wave equation [19]. This distinguishes our method from other techniques such as scalar or vector diffraction theory, which achieves scalability through approximations that are typically valid only when the characteristic unit-cell size is much larger than the wavelength [32]. As the minimum feature sizes used in our study ($\lambda/6$ for Figs. 3 and 4) violate this condition, our results reveal physical effects that cannot be captured by these approximation-based theories. For example, scalar diffraction theory neglects polarization effects entirely, which cannot accurately model the polarization-dependent anisotropy of the 3D PSF. Additionally, while the Fresnel lens we used as a baseline relies on the thin transparency approximation in diffraction theory [37], our results at high-NA lens optimization show this approach leads to suboptimal performance. This discrepancy confirms the necessity of an efficient full-wave inverse design approach for high-NA diffractive lenses at larger scales.

## 5. Conclusion

In summary, we demonstrate large-scale inverse photonic design of a diffractive lens, achieving substantial gains in focusing efficiency and resolution using only consumer-grade hardware. We anticipate future applications in 3D-printed fiber-tip lenses for optical trapping, where the enhanced focal intensity directly benefits quantum-limited control and detection of levitated nanoparticles [7]. Beyond trapping, this fast, large-scale inverse-design framework can be readily extended to a variety of industrial applications, such as free-form mask designs over a large volume [22,31], compact high-throughput microscopy [1], and laser-writing systems [3]. We also expect the deployment of the proposed method to various linear photonic devices, such as wide-field achromatic [38] and polarization-sensitive [39] metalenses, mode-selective waveguides [40], and high-Q nanocavities [41].

**Funding.** This research was supported by Ministry of Science, Research and Arts of Baden-Württemberg, Germany. M.L. acknowledges the support funded by the Alexander von Humboldt Foundation.

**Acknowledgment.** We thank Dohyeon Lee, KAIST, for their valuable discussions.

**Disclosures.** The authors declare no conflicts of interest.

**Data Availability Statement.** Data underlying the results presented in this paper are available upon request.

**Supplemental Document.** See Supplement 1 for supporting content.

### References

1. T. Gissibl, S. Thiele, A. Herkommer, and H. Giessen, "Two-photon direct laser writing of ultracompact multi-lens objectives," Nat. Photon. **10**, 554–560 (2016).
2. X. Ni, A. V. Kildishev, and V. M. Shalaev, "Metasurface holograms for visible light," Nat. Commun. **4**, 2807 (2013).

3. W. Hadibrata, H. Wei, S. Krishnaswamy, and K. Aydin, "Inverse Design and 3D Printing of a Metalens on an Optical Fiber Tip for Direct Laser Lithography," Nano Lett. **21**, 2422–2428 (2021).
4. M. Lee, H. Hugonnet, M. J. Lee, Y. Cho, and Y. Park, "Optical trapping with holographically structured light for single-cell studies," Biophysics Reviews **4**, (2023).
5. W. D. Phillips, "Nobel Lecture: Laser cooling and trapping of neutral atoms," Rev. Mod. Phys. **70**, 721–741 (1998).
6. U. Delić, M. Reisenbauer, K. Dare, D. Grass, V. Vuletić, N. Kiesel, and M. Aspelmeyer, "Cooling of a levitated nanoparticle to the motional quantum ground state," Science **367**, 892–895 (2020).
7. L. Magrini, P. Rosenzweig, C. Bach, A. Deutschmann-Olek, S. G. Hofer, S. Hong, N. Kiesel, A. Kugi, and M. Aspelmeyer, "Real-time optimal quantum control of mechanical motion at room temperature," Nature **595**, 373–377 (2021).
8. C. Gonzalez-Ballestero, M. Aspelmeyer, L. Novotny, R. Quidant, and O. Romero-Isart, "Levitodynamics: Levitation and control of microscopic objects in vacuum," Science **374**, eabg3027 (2021).
9. K. Shen, Y. Duan, P. Ju, Z. Xu, X. Chen, L. Zhang, J. Ahn, X. Ni, and T. Li, "On-chip optical levitation with a metalens in vacuum," Optica **8**, 1359–1362 (2021).
10. A. Asadollahbaik, S. Thiele, K. Weber, A. Kumar, J. Drozella, F. Sterl, A. M. Herkommer, H. Giessen, and J. Fick, "Highly Efficient Dual-Fiber Optical Trapping with 3D Printed Diffractive Fresnel Lenses," ACS Photonics **7**, 88–97 (2020).
11. S. K. Alavi, M. M. Romero, P. Ruchka, S. Jakovljević, H. Giessen, and S. Hong, "Compact vacuum levitation and control platform with a single 3D-printed fiber lens," arXiv 2504.15734 (2025).
12. M. Khorasaninejad, W. T. Chen, R. C. Devlin, J. Oh, A. Y. Zhu, and F. Capasso, "Metalenses at visible wavelengths: Diffraction-limited focusing and subwavelength resolution imaging," Science **352**, 1190–1194 (2016).
13. D. Sang, M. Xu, M. Pu, F. Zhang, Y. Guo, X. Li, X. Ma, Y. Fu, and X. Luo, "Toward High-Efficiency Ultrahigh Numerical Aperture Freeform Metalens: From Vector Diffraction Theory to Topology Optimization," Laser Photonics Rev. **16**, 2200265 (2022).
14. A. Majumder, J. A. Doughty, T. H. H. H. I. Smith, and R. Menon, "Contrary to widespread belief, the Fresnel zone plate outperforms the metalens at high NA," arXiv 2502.20481 (2025).
15. J. Jiang, M. Chen, and J. A. Fan, "Deep neural networks for the evaluation and design of photonic devices," Nat. Rev. Mater. **6**, 679–700 (2021).
16. S. Molesky, Z. Lin, A. Y. Piggott, W. Jin, J. Vucković, and A. W. Rodriguez, "Inverse design in nanophotonics," Nature Photonics **12**, 659–670 (2018).
17. N. Aage, E. Andreassen, B. S. Lazarov, and O. Sigmund, "Giga-voxel computational morphogenesis for structural design," Nature **550**, 84–86 (2017).
18. G. Osnabrugge, S. Leedumrongwatthanakun, and I. M. Vellekoop, "A convergent Born series for solving the inhomogeneous Helmholtz equation in arbitrarily large media," J. Comput. Phys. **322**, 113–124 (2016).
19. B. Krüger, T. Brenner, and A. Kienle, "Solution of the inhomogeneous Maxwell's equations using a Born series," Opt. Express **25**, 25165–25182 (2017).
20. M. Lee, H. Hugonnet, and Y. Park, "Inverse problem solver for multiple light scattering using modified Born series," Optica **9**, 177–182 (2022).
21. H. Hugonnet, M. Lee, S. Shin, and Y. Park, "Vectorial inverse scattering for dielectric tensor tomography: overcoming challenges of reconstruction of highly scattering birefringent samples," Opt. Express **31**, 29654–29663 (2023).
22. P. He, J. Liu, H. Gu, H. Jiang, and S. Liu, "Linearized EUV mask optimization based on the adjoint method," Opt. Express **32**, 8415–8424 (2024).
23. A. F. Oskooi, D. Roundy, M. Ibanescu, P. Bermel, J. D. Joannopoulos, and S. G. Johnson, "MEEP: A flexible free-software package for electromagnetic simulations by the FDTD method," Comput. Phys. Commun. **181**, 687–702 (2010).
24. W. Shin and S. Fan, "Choice of the perfectly matched layer boundary condition for frequency-domain Maxwell's equations solvers," J. Comput. Phys. **231**, 3406–3431 (2012).
25. R. E. Christiansen and O. Sigmund, "Inverse design in photonics by topology optimization: tutorial," J. Opt. Soc. Am. B **38**, 496–509 (2021).
26. S. Dottermusch, D. Busko, M. Langenhorst, U. W. Paetzold, and B. S. Richards, "Exposure-dependent refractive index of Nanoscribe IP-Dip photoresist layers," Opt. Lett. **44**, 29–32 (2018).
27. D. P. Kingma and J. Ba, "Adam: A Method for Stochastic Optimization," arXiv 1412.6980v9 (2014).
28. D. E. Chang, C. A. Regal, S. B. Papp, D. J. Wilson, J. Ye, O. Painter, H. J. Kimble, and P. Zoller, "Cavity opto-mechanics using an optically levitated nanosphere," Proc Natl Acad Sci U S A **107**, 1005–1010 (2010).
29. O. Romero-Isart, A. C. Pflanzer, F. Blaser, R. Kaltenbaek, N. Kiesel, M. Aspelmeyer, and J. I. Cirac, "Large Quantum Superpositions and Interference of Massive Nanometer-Sized Objects," Phys. Rev. Lett. **107**, 020405 (2011).
30. M. Lee, T. Hanke, S. Launer, and S. Hong, "Modelling optomechanical responses in optical tweezers beyond paraxial limits," Sci. Rep. **15**, 19377 (2025).
31. D. Lee, M. Lee, B. Yerenzhep, M. Kim, H. Hugonnet, S. Jeon, J. Shin, and Y. Park, "Fast Free-Form Phase Mask Design for Three-Dimensional Photolithography Using Convergent Born Series," ACS Photonics **12**, 610–619 (2025).

32. R. Pestourie, C. Pérez-Arancibia, Z. Lin, W. Shin, F. Capasso, and S. G. Johnson, "Inverse design of large-area metasurfaces," Opt. Express, OE **26**, 33732–33747 (2018).
33. R. Pestourie, W. Yao, B. Kanté, and S. G. Johnson, "Efficient Inverse Design of Large-Area Metasurfaces for Incoherent Light," ACS Photonics **10**, 854–860 (2023).
34. R. E. Christiansen, Z. Lin, C. Roques-Carmes, Y. Salamin, S. E. Kooi, J. D. Joannopoulos, M. Soljačić, and S. G. Johnson, "Fullwave Maxwell inverse design of axisymmetric, tunable, and multi-scale multi-wavelength metalenses," Opt. Express, OE **28**, 33854–33868 (2020).
35. M. Choi, J. Kim, S. Moon, K. Shin, S.-W. Nam, Y. Park, D. Kang, G. Jeon, K. Lee, D. H. Yoon, Y. Jeong, C.-K. Lee, and J. Rho, "Roll-to-plate printable RGB achromatic metalens for wide-field-of-view holographic near-eye displays," Nat. Mater. **24**, 535–543 (2025).
36. M. Sun, A. Shakeri, A. Keshvari, D. Giannakopoulos, Q. Wang, W.-T. Chen, S. G. Johnson, and Z. Lin, "Scalable Freeform Optimization of Wide-Aperture 3D Metalenses by Zoned Discrete Axisymmetry," ACS Photonics **12**, 3163–3171 (2025).
37. M. E. Kandel, K. W. Teng, P. R. Selvin, and G. Popescu, "Label-Free Imaging of Single Microtubule Dynamics Using Spatial Light Interference Microscopy," ACS Nano **11**, 647–655 (2017).
38. F. Aieta, M. A. Kats, P. Genevet, and F. Capasso, "Multiwavelength achromatic metasurfaces by dispersive phase compensation," Science **347**, 1342–1345 (2015).
39. N. A. Rubin, G. D'Aversa, P. Chevalier, Z. Shi, W. T. Chen, and F. Capasso, "Matrix Fourier optics enables a compact full-Stokes polarization camera," Science **365**, eaax1839 (2019).
40. Y. Meng, Y. Chen, L. Lu, Y. Ding, A. Cusano, J. A. Fan, Q. Hu, K. Wang, Z. Xie, Z. Liu, Y. Yang, Q. Liu, M. Gong, Q. Xiao, S. Sun, M. Zhang, X. Yuan, and X. Ni, "Optical meta-waveguides for integrated photonics and beyond," Light Sci. Appl. **10**, 235 (2021).
41. G. H. Ahn, K. Y. Yang, R. Trivedi, A. D. White, L. Su, J. Skarda, and J. Vučković, "Photonic Inverse Design of On-Chip Microresonators," ACS Photonics **9**, 1875–1881 (2022).

## EFFICIENT, INVERSE LARGE-SCALE OPTIMIZATION OF DIFFRACTIVE LENSES: SUPPLEMENTAL DOCUMENT

### S1. Forward solver

We present the mathematical formulation of our forward solver, which utilizes the convergent Born series (CBS) to solve the inhomogeneous wave equation for a monochromatic light field.

The total electric field $\mathbf{E}(\mathbf{r})$ satisfies the following equation:

$$\nabla \times [\nabla \times \mathbf{E}(\mathbf{r})] - k^2[1 + \chi_e S(\mathbf{r})]\mathbf{E}(\mathbf{r}) = \mathbf{0}, \tag{S1}$$

where $k = 2\pi/\lambda$ is the wavenumber, and $\chi_e = n^2 - 1 = 1.37$ is the susceptibility of the lens. The 3D lens shape is represented by a shape function $S(\mathbf{r}) = \Theta(z(h(\mathbf{r}_\perp) - z))$, where $h(\mathbf{r}_\perp)$ is the 2D thickness map and $\Theta(x)$ is a Heaviside step function. In the presence of the incident field $\mathrm{E}_{in}(\mathbf{r})$, the total electric field satisfies the following integral equation:

$$\mathbf{E}(\mathbf{r}) = \mathbf{E}_{in}(\mathbf{r}) + k^2\chi_e \int \overleftrightarrow{\mathbf{G}}(\mathbf{r} - \mathbf{r}') \cdot S(\mathbf{r}')\mathbf{E}(\mathbf{r}')d^3\mathbf{r}', \tag{S2}$$

where $\overleftrightarrow{\mathbf{G}}(\mathbf{r}) = \left[\mathbf{I}_3 + \frac{1}{k^2}\nabla\nabla^T\right]\frac{e^{ikr}}{4\pi r}$ is the dyadic Green's tensor, $r = |\mathbf{r}|$, and $\mathbf{I}_n$ is the $n \times n$ identity matrix.

For numerical computation, we discretize the fields and the shape function, representing them as vectors, such that $\mathbf{E}(\mathbf{r}) \equiv \mathbf{e} = \left[\mathbf{e}_x^T, \mathbf{e}_y^T, \mathbf{e}_z^T\right]^T$ ($3N \times 1$ vector, where $N$ represents the voxel number and 3 arises from $x$-, $y$-, and $z$-components) and $S(\mathbf{r}) \equiv \mathbf{s}$ ($N \times 1$ vector). The integral equation is then expressed as a linear matrix equation:

$$\mathbf{e} = \mathbf{e}_{in} + k^2\chi_e\mathbf{G}[\mathbf{I}_3 \otimes \mathrm{diag}(\mathbf{s})]\mathbf{e}, \tag{S3}$$

where $\otimes$ is the Kronecker product, $\mathrm{diag}(\mathbf{x})$ is diagonalized matrix of a vector $\mathbf{x}$, and $\mathbf{G} = [\mathbf{G}_{1\le i\le 3, 1\le j\le 3}]$ is a $3N \times 3N$ matrix representation of the dyadic Green's tensor. Here, $\mathbf{G}_{ij} = \mathbf{U}^\dagger\left[\left(\delta_{ij} - \frac{\mathbf{q}_i\mathbf{q}_j^T}{k^2}\right)\frac{1}{|\mathbf{q}|^2 - k^2}\right]\mathbf{U}$ is a $N \times N$ $(i, j)$-th block matrix component of $\mathbf{G}$, $\delta_{ij}$ is the Kronecker delta, and $\mathbf{U}$ represents the unitary Fourier transform operation mapping a $N \times 1$ vector in $N$ real-space 3D discrete coordinate points $\mathbf{r} = (\mathbf{x}, \mathbf{y}, \mathbf{z})$ to $N$ Fourier-space 3D discrete coordinate points $\mathbf{q} = (\mathbf{q}_x, \mathbf{q}_y, \mathbf{q}_z)$. The solution to Eq. (S3) is expressed as matrix inversion to compute $\mathbf{e}$ from given $\mathbf{e}_{in}$ and $\mathbf{s}$ [1]:

$$\mathbf{e} = \left[\mathbf{I}_{3N} - k^2\chi_e\mathbf{G}[\mathbf{I}_3 \otimes \mathrm{diag}(\mathbf{s})]\right]^{-1}\mathbf{e}_{in}. \tag{S4}$$

Since direct matrix inversion is computationally inefficient for large problems, the conventional Born series solves the equation with a Taylor series expansion:

$$\mathbf{e} = \textstyle\sum_{m=0}^{\infty}\{k^2\chi_e\mathbf{G}[\mathbf{I}_3 \otimes \mathrm{diag}(\mathbf{s})]\}^m\,\mathbf{e}_{in}. \tag{S5}$$

However, this generally diverges as the spectral radius of $k^2\chi_e\mathbf{G}[\mathbf{I}_3 \otimes \mathrm{diag}(\mathbf{s})]$ is typically outside the radius of convergence. The CBS method resolves this by modifying the Taylor series to a convergent form [2]. The modified Born series is expressed as [3]:

$$\mathbf{e} = \mathbf{e}_{in} + \frac{ik^2\chi_e}{\eta}\textstyle\sum_{m=0}^{\infty}\{\mathbf{M}(\mathbf{s})\}^m\,[\mathbf{I}_3 \otimes \mathrm{diag}(k^2\chi_e\mathbf{s} - i\eta)]\mathbf{G}_\eta[\mathbf{I}_3 \otimes \mathrm{diag}(\mathbf{s})]\mathbf{e}_{in}, \tag{S6}$$

where $\mathbf{M}(\mathbf{s}) = \mathbf{I}_{3N} + i[\mathbf{I}_3 \otimes \mathrm{diag}(k^2\chi_e \mathbf{s} - i\eta)]\{\mathbf{G}_\eta[\mathbf{I}_3 \otimes \mathrm{diag}(k^2\chi_e \mathbf{s} - i\eta)] - \mathbf{I}_{3N}\}/\eta$ . $\mathbf{G}_\eta = [\mathbf{G}_{\eta,1\le i\le 3,1\le j\le 3}]$ is a modified dyadic Green's tensor, with (*i, j*)-th block matrix component represented as $\mathbf{G}_{\eta,ij} = \mathbf{U}^\dagger\left[\left(\delta_{ij} - \frac{\mathbf{q}_i\mathbf{q}_j^T}{k^2}\right)\frac{1}{|\mathbf{q}|^2 - k^2 - i\eta}\right]\mathbf{U}$ . $\eta$ is an arbitrary constant which guarantees the convergence of the series to a solution when $\eta$ is larger than $k^2\chi_e$. The numerical computation of Eq. (S6) can efficiently be performed on a GPU by utilizing FFT3 algorithm [4].

### S2. Optimization problem and gradient calculation

Our objective functional is the intensity at a focal position $\mathbf{r}_0$, which can be represented in a discrete space as $\mathcal{L}[\mathbf{e}, \mathbf{e}^*] = \mathbf{e}^\dagger \mathbf{P}\mathbf{e}$. Here, $\mathbf{P} = \mathbf{I}_3 \otimes \mathrm{diag}(\delta(\mathbf{r} - \mathbf{r}_0))$ is a projection operator that calculates the light intensity at $\mathbf{r} = \mathbf{r}_0$. The derivative of the objective functional with respect to *m*-th component of the shape function vector, $s_m$, is derived using the chain rule:

$$\frac{\delta \mathcal{L}}{\delta s_m} = \frac{\partial \mathcal{L}}{\partial s_m} + \frac{\partial \mathcal{L}}{\partial \boldsymbol{\psi}} \cdot \frac{\delta \boldsymbol{\psi}}{\delta s_m} = \frac{\partial \mathcal{L}}{\partial \boldsymbol{\psi}} \cdot \frac{\delta \boldsymbol{\psi}}{\delta s_m}, \tag{S7}$$

where $\boldsymbol{\psi} = [\mathbf{e}^T, \mathbf{e}^\dagger]^T$ is the state variable of the fields, dependent upon $\mathbf{s}$. To efficiently compute the functional derivative $\delta\boldsymbol{\psi}/\delta s_m$, we introduce an auxiliary functional, $\mathbf{F}[\mathbf{e}, \mathbf{e}^*, \mathbf{s}] = \left[(\hat{\mathbf{f}}[\mathbf{s}]\mathbf{e})^T, \mathrm{c.c.}\right]^T = \mathbf{0}$, where $\hat{\mathbf{f}}[\mathbf{s}]\mathbf{e}$ is the discrete form of the inhomogeneous wave equation (Eq. S1):

$$\hat{\mathbf{f}}[\mathbf{s}]\mathbf{e} = \nabla \times [\nabla \times \mathbf{e}] - k^2[\mathbf{I}_3 \otimes \mathrm{diag}(1 + \chi_e \mathbf{s})]\mathbf{e} = \mathbf{0}. \tag{S8}$$

The derivative of $\mathbf{F}[\mathbf{e}, \mathbf{e}^*, \mathbf{s}]$ satisfies the following relation:

$$\frac{\delta \mathbf{F}}{\delta s_m} = \frac{\partial \mathbf{F}}{\partial s_m} + \frac{\partial \mathbf{F}}{\partial \boldsymbol{\psi}} \cdot \frac{\delta \boldsymbol{\psi}}{\delta s_m} = \mathbf{0}, \tag{S9}$$

where $\partial\mathbf{F}/\partial\boldsymbol{\psi} = \begin{bmatrix} \partial(\hat{\mathbf{f}}[\mathbf{s}]\mathbf{e})/\partial\mathbf{e} & \partial(\hat{\mathbf{f}}[\mathbf{s}]\mathbf{e})/\partial\mathbf{e}^* \\ \partial(\hat{\mathbf{f}}[\mathbf{s}]\mathbf{e})^*/\partial\mathbf{e} & \partial(\hat{\mathbf{f}}[\mathbf{s}]\mathbf{e})^*/\partial\mathbf{e}^* \end{bmatrix} = \begin{bmatrix} \hat{\mathbf{f}}[\mathbf{s}] & \mathbf{0} \\ \mathbf{0} & \hat{\mathbf{f}}[\mathbf{s}] \end{bmatrix}$ is a 2 × 2 block matrix, defined according to the convention of matrix calculus. Inserting Eq. S9 to Eq. S7, we substitute the functional derivative with partial gradients:

$$\frac{\delta \mathcal{L}}{\delta s_m} = -\underbrace{\frac{\partial \mathcal{L}}{\partial \boldsymbol{\psi}} \cdot \left(\frac{\partial \mathbf{F}}{\partial \boldsymbol{\psi}}\right)^{-1}}_{=[\mathbf{a}^T,\mathrm{c.c.}]} \cdot \frac{\partial \mathbf{F}}{\partial s_m} = -2\mathcal{R}\left[\mathbf{a}^T \cdot \frac{\partial}{\partial s_m}(\hat{\mathbf{f}}[\mathbf{s}]\mathbf{e})\right] = 2k^2\chi_e \mathcal{R}(\mathbf{a}_m^T \cdot \mathbf{e}_m), \tag{S10}$$

where $\mathbf{a}$ is the adjoint field satisfying the following wave equation:

$$\hat{\mathbf{f}}[\mathbf{s}]\mathbf{a} = \nabla \times [\nabla \times \mathbf{a}] - k^2[\mathbf{I}_3 \otimes \mathrm{diag}(1 + \chi_e \mathbf{s})]\mathbf{a} = \left(\frac{\partial \mathcal{L}}{\partial \mathbf{e}}\right)^T = \mathrm{P}\mathbf{e}^* = \mathbf{e}^*(\mathbf{r}_0). \tag{S11}$$

Therefore, computing the gradient requires two forward-solver steps: one to compute $\mathbf{e}$ and a second to compute $\mathbf{a}$.

### S3. Shape constraint

To limit the design freedom of $S(\mathbf{r})$ to a manufacturable 2D thickness profile $h(\mathbf{r}_\perp)$, we introduce a set of shape constraints similar to previous work [5]. In the discrete space, we represent $h(\mathbf{r}_\perp)$ as a $N \times 1$ discrete vector $\mathbf{h}$ satisfying $\mathrm{h}(x_m, y_m, z_m) = h(x_m, y_m)$, which is mapped to the 3D lens shape $\mathbf{s}$ through interpolation functions:

$$\mathbf{s} = \Theta\big(\mathbf{z}(\mathbf{h} - \mathbf{z})\big), \quad \text{(2D to 3D mapping)}$$

$$\begin{aligned} \mathbf{h} &= \max\big(\min\big(\mathbf{h}^{(2)}, h_{\max}\big), 0\big), \quad \text{(range limitation)} \\ \mathbf{h}^{(2)} &= \textstyle\sum_p D_{pq} \mathrm{h}_q^{(1)}, \quad \text{(symmetry over } x\text{- and } y\text{-axes)} \\ \mathbf{h}_p^{(1)} &= \textstyle\sum_l C_{lp} \mathrm{h}_l^{(0)}. \quad \text{(smoothing)} \end{aligned} \tag{S12}$$

Here, $D_{pq}$ represents the linear mirror symmetry operator along $x$- and $y$-axes. $C_{lp}$ represents a Gaussian smoothing operator. $\mathbf{h}^{(n)} \equiv B^{(n)}(\mathbf{r}_\perp)$ denotes a 2D interpolation function that defines the intermediate thickness map of the lens. The gradient of the objective function satisfies $\delta\mathcal{L}/\delta \mathrm{h}_l^{(0)} = \sum_m \big(ds_m/d\mathrm{h}_l^{(0)}\big)(\partial\mathcal{L}/\delta s_m)$, where:

$$\frac{ds_m}{d\mathrm{h}_l^{(0)}} = \textstyle\sum_p C_{lp} D_{pm} \Theta[\mathrm{h}_m(h_{\max} - \mathrm{h}_m)]\Theta(z_m)\delta(\mathrm{h}_m - z_m). \tag{S13}$$

In the discrete space, the Dirac delta $\delta(\mathrm{h}_m - z_m)$ was approximated using the subpixel smoothing method, which distributes the contribution of the interface across nearby pixels [6].

## S4. Optical trapping model

In the Rayleigh regime, the angular trapping frequencies ($\Omega_j$) of optically trapped nanoparticles for a general elliptically polarized light can be determined by the following equation [7]:

$$\Omega_j = \sqrt{\frac{6I_0}{c\rho}\left(\frac{n^2-1}{n^2+2}\right)\frac{1}{w_j}}. \tag{S14}$$

Here, $I_0$ is the focal intensity of the trapping beam, $\rho$ is the particle's density, and $w_j$ is the beam width parameter of the 3D point spread function (PSF) along the $j$-th axis, which is fitted with a Gaussian distribution, $I(\mathbf{r}) \approx I_0 \exp\left[-2\left(\frac{x^2}{w_x^2} + \frac{y^2}{w_y^2} + \frac{z^2}{w_z^2}\right)\right]$. The relationship between $w_j$ and the full-width half maxima satisfies $\mathrm{FWHM}_j = \sqrt{2\ln 2}\, w_j$. The focal power, $P_f$, is proportional to the incident laser power, $P_0$, such that $P_f = \eta_f P_0$, where $\eta_f$ is the focusing efficiency. With this Gaussian approximation, $\Omega_j^2$ is proportional to $\eta_f/(w_x w_y w_j^2)$:

$$\Omega_j \approx \sqrt{\frac{12\eta_f P_0}{c\rho w_x w_y}\left(\frac{n^2-1}{n^2+2}\right)\frac{1}{w_j}}. \tag{S15}$$

## References

1. M. Lee, H. Hugonnet, and Y. Park, "Inverse problem solver for multiple light scattering using modified Born series," Optica **9**(2), 177–182 (2022).
2. G. Osnabrugge, S. Leedumrongwatthanakun, and I. M. Vellekoop, "A convergent Born series for solving the inhomogeneous Helmholtz equation in arbitrarily large media," J. Comput. Phys. **322**, 113–124 (2016).
3. B. Krüger, T. Brenner, and A. Kienle, "Solution of the inhomogeneous Maxwell's equations using a Born series," Opt. Express **25**(21), 25165–25182 (2017).
4. M. Frigo and S. G. Johnson, "The Design and Implementation of FFTW3," Proceedings of the IEEE **93**(2), 216–231 (2005).
5. R. E. Christiansen and O. Sigmund, "Inverse design in photonics by topology optimization: tutorial," J. Opt. Soc. Am. B **38**(2), 496–509 (2021).

6. H. Foroosh, J. B. Zerubia, and M. Berthod, "Extension of phase correlation to subpixel registration," IEEE transactions on image processing **11**(3), 188–200 (2002).
7. M. Lee, T. Hanke, S. Launer, and S. Hong, "Modelling optomechanical responses in optical tweezers beyond paraxial limits," Sci. Rep. **15**(1), 19377 (2025).